\documentclass[12pt]{article}

\usepackage{amsthm,graphicx,cite}
\usepackage{epsf,latexsym}
\usepackage{amssymb,amsmath}

\newcommand{\rmi}{{\rm i}}
\newcommand{\rme}{{\rm e}}
\newcommand{\rmd}{{\rm d}}
\newcommand{\gs}{|z_{\text{\tiny R}}\rangle}
\newcommand{\gn}{|z_{n}\rangle}
\newcommand{\zr}{z_{\text{\tiny R}}}
\newcommand{\gr}{\Gamma _{\text{\tiny R}}}

\newcommand{\er}{E_{\text{\tiny R}}}
\newcommand{\Cr}{c_{\text{\tiny R}}}
\newcommand{\Co}{c_{\text{\tiny R,1}}}
\newcommand{\Ct}{c_{\text{\tiny R,2}}}

\begin{document}

\title{The decay widths, the decay constants, \\ and the
branching fractions of a resonant state}

\author{Rafael de la Madrid \\
\small{\it Department of Physics, Lamar University,
Beaumont, TX 77710} \\
\small{E-mail: \texttt{rafael.delamadrid@lamar.edu}}}

\date{\small{May 11, 2015}}












\maketitle

\begin{abstract}
\noindent 
We introduce the differential and the total decay widths of 
a resonant (Gamow) state decaying into a continuum of stable states. When 
the resonance has several decay modes, we introduce the corresponding
partial decay widths and branching fractions. In the 
approximation that the resonance is sharp, the expressions for the 
differential, partial and total decay widths of a resonant state bear a 
close resemblance with the Golden Rule. In
such approximation, the branching fractions of a resonant state are the same
as the standard branching fractions obtained by way of the Golden Rule. We
also introduce dimensionless decay constants along with their associated 
differential decay constants, and we express experimentally measurable 
quantities such as the branching fractions and the energy distributions 
of decay events in terms of those dimensionless decay constants.
\end{abstract}



\newpage

\section{Introduction}
\setcounter{equation}{0}
\label{sec:intro}

The standard Golden Rule allows us to calculate the transition rate 
(probability of transition per unit time) from an energy eigenstate of a 
quantum system into a continuum of energy eigenstates. If 
$|E_{\rm i}\rangle$ is the eigenstate of 
an unperturbed Hamiltonian $H_0$, and if such state is coupled to
a state $|E_{\rm f}\rangle$ by a perturbation $V$, the transition 
probability per unit of time from the initial state 
$|E_{\rm i}\rangle$ to the final
state $|E_{\rm f}\rangle$ is given, to first order in the perturbation, by
\begin{equation}
      R_{{\rm i}\to {\rm f}} = \frac{2 \pi} {\hbar} 
          \left| \langle E_{\rm f}|V|E_{\rm i} \rangle \right |^{2} 
            \delta (E_{\rm i}-E_{\rm f})  \, ,
            \label{usualgr0}
\end{equation}
where $\langle E_{\rm f}|V|E_{\rm i} \rangle$ is the matrix element 
of the perturbation $V$ between the final and initial states. If 
the initial state is coupled to a continuum of final states 
$|E_{\rm f}\rangle$, and if the density of final states
(number of states per unit of energy) is $\rho (E_{\rm f})$, the transition 
probability per unit of time from the state $|E_{\rm i}\rangle$ to the
continuum of final
states $|E_{\rm f}\rangle$ is given by
\begin{equation}
      \overline{R}_{{\rm i}\to {\rm f}} = \int \rmd E_{\rm f} \, 
             \rho (E_{\rm f}) R_{{\rm i}\to {\rm f}} =
        \frac{2 \pi} {\hbar} 
          \left| \langle E_{\rm f}|V|E_{\rm i} \rangle \right |^{2} 
           \rho (E_{\rm i})  \, ,
            \label{usualgr1}
\end{equation}
where it has been assumed that the matrix element 
$\langle E_{\rm f}|V|E_{\rm i} \rangle$ is the same for all the states in
the continuum. The decay width from the initial state $|E_{\rm i}\rangle$ 
to the final state $|E_{\rm f}\rangle$ is given by
\begin{equation}
     \Gamma_{{\rm i}\to {\rm f}}=\hbar 
         R_{{\rm i}\to {\rm f}} = 
            2 \pi 
          \left| \langle E_{\rm f}|V|E_{\rm i} \rangle \right |^{2} 
            \delta (E_{\rm i}-E_{\rm f}) \, .
            \label{usualgr2}
\end{equation}
The decay width from the initial state $|E_{\rm i}\rangle$ to a continuum of 
final states $|E_{\rm f}\rangle$ is given by
\begin{equation}
     \overline{\Gamma}_{{\rm i}\to {\rm f}}=\hbar 
         \overline{R}_{{\rm i}\to {\rm f}} = 
          2 \pi 
          \left| \langle E_{\rm f}|V|E_{\rm i} \rangle \right |^{2} 
           \rho (E_{\rm i})  \, .  
            \label{usualgr3}
\end{equation}
The standard derivation of Eqs.~(\ref{usualgr0})-(\ref{usualgr3}) is 
the result of first-order perturbation theory, and it is valid when 
the initial and final states are stable. The purpose of the present 
paper is to introduce the analog
of Eqs.~(\ref{usualgr0})-(\ref{usualgr3}) under the assumption that the initial 
state is described by a resonant (Gamow) state, and that such state
decays into a continuum of stable, scattering states. 

In Sec.~\ref{sec:preliminaries}, we introduce the differential and the total
decay widths of a resonant state that has only one decay mode. In 
Sec.~\ref{sec:derivation}, we derive an expression for such decay widths 
in terms of a truncated Breit-Wigner (Lorentzian) lineshape
and the matrix element of the interaction. We also show that, when the 
resonance is sharp and far from the energy threshold, the expressions for 
the resonant decay widths bear a strong
resemblance with Fermi's Golden Rule. In Sec.~\ref{sec:application},
we apply the results of Sec.~\ref{sec:derivation} to the 
delta-shell potential. In Sec.~\ref{sec:branching}, we 
introduce the partial decay widths of a resonant state that has more than 
one decay mode, we define the branching fractions for each decay mode, 
and we point out that, at least in principle, such branching fractions 
afford a way to falsify the formalism of the present 
paper. In Sec.~\ref{sec:ddc}, we introduce dimensionless partial and total
decay constants along with their associated differential decay constants, 
we express the branching fractions in
terms of them, and we argue that the differential decay constant corresponds
to a measurement of the energy distributions of decay events (the invariant 
mass distributions 
of particle physics). Section~\ref{sec:conclusions} contains our conclusions.

\section{Preliminaries}
\setcounter{equation}{0}
\label{sec:preliminaries}

Let $H=H_0+V$ be a Hamiltonian that produces resonances, where $H_0$ is the
free Hamiltonian and $V$ is the interaction potential. The continuum spectra
of both $H$ and $H_0$ will be assumed to be $[0,\infty)$, as it often occurs
in non-relativistic quantum mechanics. For simplicity, we
will assume that the continuum spectra are non-degenerate, and that 
the eigenstates
of $H_0$ can be determined by a single quantum number, the
energy $E$. The eigenstates of the free Hamiltonian will be denoted by
$|E\rangle$,
\begin{equation}
       H_0|E\rangle = E |E\rangle \, .
\end{equation}
We will denote the resonant state by $|\zr \rangle$,
\begin{equation}
       H|\zr\rangle = \zr |\zr \rangle \, ,
\end{equation}
where $\zr = \er - \rmi \gr /2$ is the complex resonant energy of a 
decaying state. We 
will denote by $\rme ^{-\rmi H\tau /\hbar}|\zr \rangle$ the time 
evolution of the resonant state, $\tau$ being the time parameter that
appears in the Schr\"odinger equation.

When $|E\rangle$ is delta normalized and $|\zr\rangle$ is normalized
according to Zeldovich's normalization, the ``scalar product'' 
$\langle E|\rme ^{-\rmi H\tau /\hbar}|\zr \rangle$ has dimensions of 
$\left(\sqrt{\rm energy}\right)^{-1}$. Thus,
$|\langle E|\rme ^{-\rmi H\tau /\hbar}|\zr \rangle|^{2}$ has dimensions of 
$\left({\rm energy}\right)^{-1}$, and it can be interpreted as a probability
density, i.e., a probability per unit of energy. This is why we 
will identify $|\langle E|\rme ^{-\rmi H\tau /\hbar}|\zr \rangle|^{2}$ with
the probability density $\frac{\rmd P_{\tau}}{\rmd E}$ 
that the resonance has decayed into a stable particle of energy $E$ at
time $\tau$,
\begin{equation}
      \frac{\rmd P_{\tau}}{\rmd E}\equiv 
         |\langle E|\rme ^{-\rmi H\tau /\hbar}|\zr \rangle|^{2}  \, .
             \label{detdpd}
\end{equation}

As is well known, the survival probability of a resonant (Gamow) 
state abides by the exponential law,
\begin{equation}
      p_{\rm s}(\tau)= \rme ^{-\gr \tau /\hbar} \, .
          \label{ed}
\end{equation}
The initial decay rate $\dot{p}_{\rm s}(0)$ associated with the 
survival probability satisfies
\begin{equation}
      \gr =- \hbar \dot{p}_{\rm s}(0) 
        = -\hbar \left. \frac{\rmd p_{\rm s}(\tau)}{\rmd \tau}\right|_{\tau =0}
       \, .
        \label{drgs}
\end{equation}

In analogy to Eq.~(\ref{drgs}), we define the differential decay width
of a resonant state as follows:
\begin{equation}
      \frac{\rmd \overline{\Gamma} (E)}{\rmd E}  \equiv
         -\hbar \frac{\rmd  \dot{P}_{\tau=0}}{\rmd E} =
         -\hbar \frac{\rmd}{\rmd \tau}\left(
               \left.\frac{\rmd  P_{\tau}}{\rmd E}\right)
                   \right|_{\tau =0}    \, ,
             \label{diffdr}
\end{equation}
where $\frac{\rmd  \dot{P}_{\tau}}{\rmd E}$ is the time derivative (rate
of change) of $\frac{\rmd P_{\tau}}{\rmd E}$. The 
total decay width is obtained by integration over the scattering
spectrum of the free Hamiltonian,
\begin{equation}
      \overline{\Gamma} = 
         \int_0^{\infty} \rmd E \ \frac{\rmd \overline{\Gamma}(E)}{\rmd E}  \, .
         \label{totaldr}
\end{equation}
This formula is valid when the energy $E$ is the only quantum number needed
to describe the stable, asymptotic states. If such states are described by 
additional quantum numbers, we just need to sum/integrate over such quantum 
numbers. 

Three comments are in order here. First, in the standard approach,
one uses the Golden Rule to calculate the width 
$\overline{\Gamma}_{{\rm i}\to {\rm f}}$ using Eq.~(\ref{usualgr3}), and then
one assumes that $\overline{\Gamma}_{{\rm i}\to {\rm f}}$ coincides with the
width $\gr$ that arises from the pole of the $S$-matrix. However, as 
discussed in Appendix~\ref{app:rla}, the value of the decay width 
$\overline{\Gamma}$ of Eq.~(\ref{totaldr}) is in general different from 
the value of $\gr$. Thus, $\overline{\Gamma}$ should be interpreted 
as another way to characterize the strength of the interaction between 
the resonance and the continuum.\footnote{Interestingly, in particle 
physics there are sometimes two and even three definitions of the width 
of unstable particles~\cite{CASO}, and only one of them coincides with
the width that arises from the pole of the 
$S$-matrix~\cite{SIRLIN,WILLENBROCK,STUART,LEIKE,BAWI,PASSERA,BERNICHA,BH,GRASSI,BS,GP}.}

Second, because the probability in Eq.~(\ref{detdpd}) can be defined also for a
square-integrable wave function $\varphi$, we can define the differential
and the total decay widths for the transition from a state $\varphi$ into
a continuum. We can then expand $\varphi$ in 
terms of the resonant (Gamow) states and a background,
$|\varphi \rangle =\Cr |\zr \rangle + |\text{bg}\rangle$, where
$|\text{bg}\rangle$ is the background, and where it has been
assumed that the system has only one resonant state $|\zr\rangle$. The 
resulting expression will have
a resonant part associated solely with $|\zr \rangle$, and other terms 
that involve $|\text{bg}\rangle$. As it usually happens with resonant 
expansions, the term associated exclusively with $|\zr\rangle$ carries 
the contribution from the resonance per se. Such resonant contribution is 
what Eqs.~(\ref{detdpd}), (\ref{diffdr}), and (\ref{totaldr}) contain. The
other terms (which, although important, are not the focus of the present
paper) contain the non-exponential contributions to the decay
constant.

Third, the resonant decay width of Eq.~(\ref{totaldr}) depends on the 
normalization of the Gamow states. When we use Zeldovich's normalization, the 
decay width has dimensions of energy, but if we used another 
normalization, the value and possibly the dimensions of $\overline{\Gamma}$ 
would change.

Before finishing this section, it may be useful to add a few words about
the role of resonant expansions in quantum mechanics. In a resonant
expansion such as $|\varphi \rangle =\Cr |\zr \rangle + |\text{bg}\rangle$,
the resonant state $|\zr \rangle$ is supposed to carry
the resonance's contribution to the state $\varphi$ (including the 
exponential decay), and the background is supposed to carry the non-resonant
contributions (including deviations from exponential decay). When one 
calculates, for example, the survival probability using a wave function 
$\varphi _1= \Co |{\zr} \rangle + |\text{bg}_1\rangle$, one usually 
obtains an exponential decay for intermediate times and deviations from
exponential decay at short and long times. However, if we used a wave
function $\varphi _2= \Ct |\zr \rangle + |\text{bg}_2\rangle$ that is only 
slightly different from $\varphi _1$, we would obtain a slightly different
survival probability with slightly different deviations from exponential
decay. The question then arises: What is the wave function
of the resonance? Is it $\varphi _1$, $\varphi _2$, or something else? Or
even worse, do $\varphi _1$ and $\varphi _2$ represent the wave functions of
two different resonances? The Gamow-state description of resonances addresses
these questions very easily: For each pole of the $S$-matrix, one can construct
a unique Gamow state that is identified with the wave function of the 
resonance. In our example, there is only one resonance, and $\varphi _1$
and $\varphi _2$ are two different approximations of one and the same 
resonant state. Thus, by identifying the resonant (Gamow) state with the 
wave function of a resonance, one has a clear way to prescribe what is 
resonance from what is not resonance. 

Identifying the Gamow state with the resonance's wave function is very similar
to identifying a monochromatic wave with a plane wave. Both a plane wave
and a Gamow state are not square-integrable functions, but when you prepare
a square integrable wave function very close to a plane wave or a Gamow
state, you have, for all purposes and intends, a monochromatic state or
a Gamow state.

\section{Derivation}
\setcounter{equation}{0}
\label{sec:derivation}

Let us assume that our system has been prepared
in a resonant state $|\zr \rangle$. Such state satisfies
the following integral equation~\cite{WOLF,MONDRA,JMP13,ARCOS}:
\begin{equation}
    \gs = \frac{1}{\zr -H_0 +\rmi 0}V \gs \, . 
         \label{inteGam0}
\end{equation}  
By multiplying Eq.~(\ref{inteGam0}) by $\rme ^{-\rmi \zr \tau/\hbar}$, and by 
taking into account that 
$\rme ^{-\rmi \zr \tau/\hbar}\gs=\rme^{-\rmi H\tau /\hbar}\gs$, we obtain that
\begin{equation}
    \rme^{-\rmi H\tau /\hbar} \gs =
      \rme^{-\rmi \zr \tau /\hbar} \frac{1}{\zr -H_0 +\rmi 0}V \gs \, . 
         \label{inteGam1}
\end{equation}
The ``scalar product'' of Eq.~(\ref{inteGam1}) with $\langle E|$ can
be written as
\begin{equation}
   \langle E| \rme^{-\rmi H\tau /\hbar} \gs =
      \rme^{-\rmi \zr \tau /\hbar} \langle E|\frac{1}{\zr -H_0 +\rmi 0}V \gs \, . 
         \label{inteGam2}
\end{equation}
When $\frac{1}{\zr -H_0 +\rmi 0}$ acts to the right, as it does
in Eq.~(\ref{inteGam2}), we have that
\begin{equation}
       \langle E|\frac{1}{\zr -H_0 +\rmi 0} = \frac{1}{\zr -E} \langle E| \, .
        \label{inter0}
\end{equation}
Substitution of Eq.~(\ref{inter0}) into Eq.~(\ref{inteGam2}) yields
\begin{equation}
   \langle E| \rme^{-\rmi H\tau /\hbar} \gs =
      \rme^{-\rmi \zr \tau /\hbar} \frac{1}{\zr -E} \langle E|V \gs \, . 
         \label{inteGam3}
\end{equation}
By combining Eqs.~(\ref{inteGam3}) and~(\ref{detdpd}), we obtain
\begin{equation}
   \frac{\rmd P_{\tau}(E)}{\rmd E}= 
       |\langle E| \rme^{-\rmi H\tau /\hbar} \gs|^2 =
      \rme^{-\gr \tau /\hbar} 
         \frac{1}{(E-\er)^2+(\gr/2)^2} |\langle E|V \gs|^2 \, . 
         \label{inteGam4}
\end{equation}
The time derivative of this expression reads
\begin{equation}
   \frac{\rmd \dot{P}_{\tau}(E)}{\rmd E} = -
      \rme^{-\gr \tau /\hbar} 
          \frac{\gr/\hbar}{(E-\er)^2+(\gr/2)^2} |\langle E|V \gs|^2 \, . 
         \label{inteGam5}
\end{equation}
Substitution of Eq.~(\ref{inteGam5}) into Eq.~(\ref{diffdr}) yields
the differential decay width of a resonant state,
\begin{equation}
   \frac{\rmd \overline{\Gamma} (E)}{\rmd E}=
       \frac{\gr}{(E-\er)^2+(\gr/2)^2} |\langle E|V \gs|^2 \, . 
         \label{inteGam6}
\end{equation}
Substitution of Eq.~(\ref{inteGam6}) into Eq.~(\ref{totaldr})
yields the total decay width of a resonant state,
\begin{equation}
           \overline{\Gamma} = \int_0^{\infty} \rmd E \
       \frac{\gr}{(E-\er)^2+(\gr/2)^2} |\langle E|V \gs|^2 \, . 
         \label{inteGam6total}
\end{equation}
Thus, the decay width of a resonant state depends on the matrix
element squared of the interaction potential and on the 
Lorentzian lineshape. The energy domain of the Lorentzian, however, does 
not extend from $-\infty$ to $+\infty$, but it is truncated at the lower energy
threshold $E=0$. Equations~(\ref{inteGam6}) 
and~(\ref{inteGam6total}) contain no 
approximations and, as we are going to see below, they bear a strong
resemblance with the standard Golden Rule of 
Eqs.~(\ref{usualgr2}) and~(\ref{usualgr3}) 
in the approximation that the resonance is sharp and far away from the 
threshold.

In order to relate Eqs.~(\ref{inteGam6}) and~(\ref{inteGam6total})
with the standard Golden Rule, we are going to use the well-known result that,
when its width tends to zero, the
Lorentzian becomes the delta function, 
\begin{equation}
    \delta(x) = \lim _{\epsilon \to 0} \frac{\epsilon/\pi}{x^2+\epsilon ^2} \, .
         \label{deltal}
\end{equation}
Let $x=(E-\er)$ and $\epsilon = \gr/2$. Then, when 
the resonance is sharp (i.e., when $\gr/(2\er)$ is small), and
when we can neglect the lower energy threshold (i.e., when the resonance 
is far from the $E=0$ threshold), we have that Eq.~(\ref{deltal}) yields
\begin{equation}
    \delta(E-\er) = \lim _{\gr \to 0} \frac{\gr/(2\pi)}{(E-\er)^2+(\gr/2) ^2}
       \, .
          \label{lgode}
\end{equation}
By combining Eqs.~(\ref{inteGam6}), (\ref{inteGam6total}) and~(\ref{lgode}), 
we obtain
\begin{eqnarray}
    && \frac{\rmd \overline{\Gamma} (E)}{\rmd E}\approx 2\pi 
         |\langle E|V \gs|^2 \delta (E-\er) \, , 
         \label{inteGam8} \\ [2ex]
    && \overline{\Gamma} \approx
          2\pi  \,  
         |\langle \er |V \gs|^2  \, . 
         \label{inteGam9}
\end{eqnarray}
Equation~(\ref{inteGam8}) has the same form as 
Eq.~(\ref{usualgr2}). Equation~(\ref{inteGam9}) has the same 
form as Eq.~(\ref{usualgr3}), except that in Eq.~(\ref{inteGam9})
the ``density of states'' is 1. 

It is surprising that the decay widths of 
Eqs.~(\ref{inteGam6}) and~(\ref{inteGam6total}) become very similar to the
standard Golden Rule in the approximation that the resonance is sharp, 
since the derivation leading to Eqs.~(\ref{inteGam8}) and~(\ref{inteGam8})
is completely different from the derivation leading to the standard 
Golden Rule. Although it is not clear why such is the case, 
in Appendix~\ref{app:ana} we explore a possible explanation.

In the literature, one can find several expressions for the transition 
probability of Eq.~(\ref{inteGam4}) and the decay widths
of Eqs.~(\ref{inteGam6}) and~(\ref{inteGam6total}), see 
Refs.~\cite{GOLDBERGER,MERZBACHER,BOHM}. Such expressions 
always involve
the matrix element squared of the interaction potential and the Lorentzian
lineshape. In particular, the decay widths~(\ref{inteGam6}) 
and~(\ref{inteGam6total}) coincide with those provided in Ref.~\cite{BOHM},
although the expressions of Ref.~\cite{BOHM} are derived from a
time-dependent probability that is different from that in 
Eq.~(\ref{inteGam4}).\footnote{Further variations 
on Fermi's Golden Rule can be found 
in Refs.~\cite{FONDA,LAMB,COSTIN,VOS,TOBITA,BRYANT,DEBIERRE,DEBIERRE2}.}

\section{An illustrative example: The delta potential}
\setcounter{equation}{0}
\label{sec:application}

We are going to use Eqs.~(\ref{inteGam6}) and~(\ref{inteGam6total}) to 
calculate the decay width of a particle trapped by a delta 
potential~\cite{WINTER,DICUS,GASTON,SANTINI1,SANTINI2} located at $r=a$,
\begin{equation}
       V(r)=g\delta (r-a) \, ,
	\label{deltap}
\end{equation}
where $g$ accounts for the strength of the potential. Since this potential
is spherically symmetric, we can work in the radial, position 
representation. For simplicity, we will restrict ourselves to s-waves
(zero angular momentum), since the higher-order cases are treated 
analogously. The eigenfunctions $\chi _0(r;E)$ of the free Hamiltonian $H_0$ 
satisfy
\begin{equation}
	-\frac{\hbar^2}{2m} \frac{\rmd ^2}{\rmd r^2}\chi_0(r;E)=
         E\chi _0(r;E)\,.
	\label{rSe0}
\end{equation}
The delta-normalized solution of Eq.~(\ref{rSe0}) is well known (see
for example Ref.~\cite{IJTP}),
\begin{equation}
      \chi _0(r;E)=\sqrt{ \frac{1}{\pi}\, \frac{2m/\hbar ^2}{k}} \,
      \sin \left(k r \right) , 
      \quad 0<r<\infty \, ,
      \label{0chi}
\end{equation}
where $k=\sqrt{\frac{2m}{\hbar ^2}E}$ is the wave number. 
The eigenfunctions $\chi (r;E)$ of the Hamiltonian $H$ 
satisfy
\begin{equation}
	\left( -\frac{\hbar^2}{2m} \frac{\rmd ^2}{\rmd r ^2} + 
             g\delta (r-a) \right)
           \chi(r;E)=
         E\chi (r;E) \, ,  
	\label{rSe}
\end{equation}
subject to the boundary conditions
\begin{eqnarray}
    &&\chi(0;E)=0 \, ,  \label{bc0} \\
    &&\chi (a+;E)= \chi (a-;E) \, ,  \label{bc1}  \\
    &&\chi '(a+;E)- \chi '(a-;E)= \frac{2mg}{\hbar ^2} \chi(a;E)\, . \label{bc2}
\end{eqnarray}
The solution of Eq.~(\ref{rSe}) subject to Eqs.~(\ref{bc0})-(\ref{bc2}) 
is given by
\begin{equation}
       \chi (r;E)=A(E) \left\{ \begin{array}{ll}
        \sin (kr) \quad  & 0<r<a \, , \\ [2ex]
        {\cal J}_1(E) \rme ^{\rmi kr} + {\cal J}_2(E) \rme ^{-\rmi kr}        
     \quad &  a<r<\infty \, , 
        \end{array} 
      \right. 
	\label{0-}
\end{equation}
where 
\begin{eqnarray}
       && {\cal J}_1(E) = \frac{\rme ^{-\rmi ka}}{2}
          \left[ \left(1-\rmi \frac{2mg}{\hbar ^2k} \right) \sin (ka)
                 -\rmi \cos (ka) \right] \, , 
               \label{j1}  \\ [2ex]
      && {\cal J}_2(E) = \frac{\rme ^{\rmi ka}}{2}
          \left[ \left(1+\rmi \frac{2mg}{\hbar ^2k} \right) \sin (ka)
                 +\rmi \cos (ka) \right]  \, ,
 	\label{j2}
\end{eqnarray}
and where $A(E)$ is an overall normalization constant. The $S$-matrix
reads
\begin{equation}
      S(E)=-\frac{ {\cal J}_1(E)}{{\cal J}_2(E)} \, .
\end{equation}

The resonant states
are obtained by imposing the purely outgoing boundary condition on the
Schr\"odinger equation. In our case, this is tantamount to imposing
that ${\cal J}_2(E)=0$,
\begin{equation}
   \left(1+\rmi \frac{2mg}{\hbar ^2k} \right) \sin (ka)
                 +\rmi \cos (ka) =0 \, .
 	\label{rc}
\end{equation}
We will denote the complex solutions of Eq.~(\ref{rc}) by 
$z_n$, $n=1,2,\ldots$. The corresponding resonant states read as follows:
\begin{equation}
       u(r;z_n)= N_n \left\{ \begin{array}{ll}
        \frac{1}{{\cal J}_1(z_n)} \sin (k_n r) \quad  & 0<r<a \, , \\ [2ex]
                  \rme ^{\rmi k_nr}        
     \quad &  a<r<\infty \, ,
        \end{array} 
      \right. 
	\label{rstate}
\end{equation}
where Zeldovich's normalization constant $N_n$ is given by the
residue of the $S$-matrix at the complex resonant wave number $k_n$, 
$N_n^2= \rmi \, \text{res}\left[S(q)\right]_{q=k_n}$.

The matrix element of the potential is given by
\begin{eqnarray}
     \langle E|V\gn  &=& \int_0^\infty \rmd r \, 
                  \overline{\chi_0(r;E)} g\delta (r-a) u(r;z_n) \nonumber \\
           &=& g\sqrt{ \frac{1}{\pi}\, \frac{2m/\hbar ^2}{k}} \,
      \sin \left(k a \right) N_n \rme^{\rmi k_n a} \, .
\end{eqnarray}
Thus, Eq.~(\ref{inteGam6}) yields
\begin{equation}
   \frac{\rmd \overline{\Gamma}_n (E)}{\rmd E}=
       \frac{\Gamma _n}{(E-E_n)^2+(\Gamma _n/2)^2} 
        \frac{2mg^2}{\pi k \hbar ^2} \,
      \sin ^2\left(k a \right) |N_n|^2 \rme^{2\beta _n a}    \, , 
         \label{inteGam6dp}
\end{equation}
where $\beta _n$ is the imaginary part of the complex wave number $k_n$, i.e.,
$k_n=\alpha _n-\rmi \beta _n$. The total decay width is given by
\begin{equation}
   \overline{\Gamma}_n = \frac{2mg^2}{\pi \hbar ^2} |N_n|^2 \rme ^{2\beta _n a} C_n
          \, ,
         \label{inteGam7dp}
\end{equation}
where
\begin{equation}
   C_n= \int_0^{\infty}\rmd E \,\frac{\Gamma _n}{(E-E_n)^2+(\Gamma _n/2)^2} 
        \frac{1}{k} \,
      \sin ^2\left(k a \right)  .
\end{equation}

Because the delta-shell potential is very simple, one can calculate the
decay width analytically. However, for more
complicated potentials, it may not be possible to calculate the resonant
states exactly, and one may have to resort to approximations.

\section{Partial decay widths and branching fractions}
\setcounter{equation}{0}
\label{sec:branching}

When a resonance has different decay modes, one usually defines the 
branching fractions to account for how likely the 
resonance will decay into each mode. Experimentally the branching fractions 
are measured as the number of particles that decay into a given mode divided 
by the total number of decays.

In order to obtain the theoretical branching fractions, we need
to define the partial decay widths. Let us assume for simplicity that the
resonance has two decay modes and that each decay mode can be described
by a single, discrete, quantum number $j=1,2$. Similarly to 
Eq.~(\ref{inteGam4}), the probability (per unit of 
energy) that the resonance decays at time $\tau$ into the mode $j$ is 
\begin{equation}
   \frac{\rmd P_{j, \tau}(E)}{\rmd E} =
       |\langle E,j| \rme^{-\rmi H\tau /\hbar} \gs|^2 =
      \rme^{-\gr \tau /\hbar} 
         \frac{1}{(E-\er)^2+(\gr/2)^2} |\langle E,j|V \gs|^2 \, , \quad
         j=1,2 \, . 
         \label{inteGam46}
\end{equation}
Similarly to Eq.~(\ref{inteGam6}), the partial differential 
decay width of a resonant state is
\begin{equation}
   \frac{\rmd \overline{\Gamma} _j(E)}{\rmd E}=
       \frac{\gr}{(E-\er)^2+(\gr/2)^2} |\langle E,j|V \gs|^2 \, , \qquad
         j=1,2 \, .
         \label{inteGam66}
\end{equation}
The partial decay width is
\begin{equation}
       \overline{\Gamma} _j = \int_0^{\infty} \rmd E \
       \frac{\gr}{(E-\er)^2+(\gr/2)^2} |\langle E,j|V \gs|^2 \, , \qquad
          j=1,2 \, . 
         \label{inteGam6total6}
\end{equation}
The resulting branching fractions are 
\begin{equation}
      {\cal B} (\text{R}\to j) = \frac{\overline{\Gamma} _j}
          {\overline{\Gamma} _1+\overline{\Gamma} _2} \, ,  
             \qquad j =1,2 \, .
            \label{br}
\end{equation}   

Similarly to Eq.~(\ref{inteGam9}), when the resonance is sharp, the partial 
decay widths are approximated by
\begin{equation}
     \overline{\Gamma} _j\approx
          2\pi  \,  
         |\langle \er, j |V \gs|^2 \, , \qquad j=1,2  \, . 
         \label{inteGam9j}
\end{equation}
In such an approximation, the branching fractions read
\begin{equation}
         {\cal B} (\text{R}\to j) \approx
\frac{\left| \langle \er ,j|V|\zr \rangle \right |^{2}}{\left| \langle \er,1|V|\zr \rangle \right |^{2}+\left| \langle \er ,2|V|\zr \rangle \right |^{2}} \, , \qquad j=1, 2 \, ,
       \label{abr}
             \end{equation}
which is the same result we would obtain if we used the decay widths
of Eq.~(\ref{usualgr3}) with $|E_{\rm i}\rangle \equiv |\zr \rangle$ and
$|E_{\rm f} \rangle \equiv |\er \rangle$. Thus, in the approximation that 
the resonance is sharp, the branching fractions obtained from the probability 
distribution of Eq.~(\ref{inteGam46}) are the same as those obtained from 
the Golden Rule. However, when the resonance is not sharp, we would obtain 
different branching fractions. Because the branching fractions do
not depend on the normalization of the resonant state, Eq.~(\ref{br}) will
in general yield a different result from that of Eq.~(\ref{abr}), no matter
what normalization is used for the resonant state. This, at least in 
principle, makes the formalism of the present paper falsifiable.

\section{The (dimensionless) decay constants}
\setcounter{equation}{0}
\label{sec:ddc}

There are two main kinds of decay experiments. In one kind, one measures the
number of decay events as a function of time, which corresponds to
measuring a time-dependent probability distribution such as the survival 
probability. In another kind, one measures the number of decay events as 
a function of the energy, and one obtains an energy distribution of decay 
events such as the invariant mass distributions of particle 
physics (see for example 
Refs.~\cite{ATLAS,CMS,LHCb,CDF,BESIII,BELLE,D0,BABAR,ALICE}). The 
purpose of this section is to describe the energy distributions of decay 
events in terms of the resonant states.

Although a scattering experiment is a time-dependent phenomenon, one usually
describes it using time-independent quantities such as the cross section and
the $S$-matrix. Such time-independent quantities can be obtained from
solutions of the time-independent Schr\"odinger equation. Similarly, a
decay process is a time-dependent phenomenon, but it can be described
by time-independent quantities such as the energy distributions of decay
events. The experimental energy distributions of decay events are simply 
the number of decay events per unit of energy as a function of the energy,
\begin{equation}
         \left( \frac{ \# \ \text{decay events}}{\text{energy}} \
           \text{vs.} \ \text{energy} \right) \, .
\end{equation}
When we divide the number of decay events by the total number of events
$N_0$, the corresponding experimental distribution corresponds to the
theoretical probability distribution of decay,
\begin{equation}
      \frac{\rmd P(E)}{\rmd E}= \left( \frac{\text{decay probability}}{\text{energy}} \
           \text{vs.} \ \text{energy}\right)
                     \ \leftrightarrow \ \left(
            \frac{1}{N_0}\frac{ \# \ \text{decay events}}{\text{energy}} \
            \text{vs.} \ \text{energy} \right) \, ,
\end{equation}
where $\frac{\rmd P (E)}{\rmd E}$ is the (theoretical) probability per unit of 
energy that the resonance has decayed into a stable particle of energy $E$.

When $|E\rangle$ is delta normalized and $|\zr\rangle$ is normalized
according to Zeldovich's normalization, the ``scalar product'' 
$\langle E|\zr \rangle$ has dimensions of 
$\left(\sqrt{\rm energy}\right)^{-1}$. Thus,
$|\langle E|\zr \rangle|^{2}$ has dimensions of 
$\left({\rm energy}\right)^{-1}$, and it is natural to identify it
with $\frac{\rmd P(E)}{\rmd E}$,
\begin{equation}
      \frac{\rmd P (E)}{\rmd E}\equiv 
         |\langle E|\zr \rangle|^{2}  \, .
             \label{detdpdti}
\end{equation}
Obviously, the probability of Eq.~(\ref{detdpdti}) is the time-independent 
version of the
probability of Eq.~(\ref{detdpd}). The total probability that the particle
has decayed into the continuum is 
\begin{equation}
      P= \int_0^{\infty} \frac{\rmd P (E)}{\rmd E}\ \rmd E \, .
             \label{detdpdtito}
\end{equation}
Similarly to the definitions of the decay widths 
$\frac{\rmd \overline{\Gamma}}{\rmd E}$ and $\overline{\Gamma}$, we can 
define the differential and the total decay constants as follows,
\begin{eqnarray}
       &\frac{\rmd \Gamma (E)}{\rmd E} \equiv \frac{\rmd P (E)}{\rmd E}
               = |\langle E|\zr \rangle|^{2} \, , 
         \label{gitp1} \\ [2ex]
       &    \Gamma \equiv P=  \
          \int_0^{\infty} |\langle E|\zr \rangle|^{2} \ \rmd E \, .
            \label{gitp2} 
\end{eqnarray}
By using Eq.~(\ref{inteGam0}), and
by following similar steps to those in Sec.~\ref{sec:derivation}, we can express
the decay constants of Eqs.~(\ref{gitp1}) and (\ref{gitp2}) as 
\begin{eqnarray}
   \frac{\rmd \Gamma (E)}{\rmd E} =
       \frac{1}{(E-\er)^2+(\gr/2)^2} |\langle E|V \gs|^2 \, , 
         \label{inteGam6ti} \\ [2ex]
           \Gamma =  \int_0^{\infty} \rmd E \
       \frac{1}{(E-\er)^2+(\gr/2)^2} |\langle E|V \gs|^2 \, . 
         \label{inteGam6totalti}
\end{eqnarray}
These expressions are very similar to the expressions for the decay
widths of Eqs.~(\ref{inteGam6}) and~(\ref{inteGam6total}). In fact, it 
follows from Eqs.~(\ref{inteGam6}), (\ref{inteGam6total}), (\ref{inteGam6ti})
and (\ref{inteGam6totalti}) that
\begin{eqnarray}
       & \frac{\rmd \Gamma (E)}{\rmd E}= \frac{1}{\gr} 
             \frac{\rmd \overline{\Gamma} (E)}{\rmd E} \, , 
            \label{rddcddw}  \\ [1ex]
      & \Gamma =  \frac{\overline{\Gamma}}{\gr} \, .
             \label{redcdw}
\end{eqnarray}
As can be seen from Eq.~(\ref{detdpdtito}) or from Eq.~(\ref{redcdw}), 
$\Gamma$ is a dimensionless constant that can be interpreted
as another way to measure the overall strength of the interaction between 
the resonance and the continuum. By contrast, $\frac{\rmd \Gamma (E)}{\rmd E}$ 
has units of $\left({\rm energy}\right)^{-1}$, and hence can be
interpreted as the probability density that the resonance decays into
a stable particle of energy $E$.

Similarly to the decay width $\overline{\Gamma}$, one can define the 
dimensionless decay constant $\Gamma$ of a square-integrable
wave function $\varphi$. When we perform a resonant expansion of $\varphi$ 
and substitute
such expansion in the expression for $\Gamma$, we obtain a contribution 
that is associated exclusively with the resonant state, and other 
contributions that involve the background. The contribution
that is associated exclusively with $|\zr \rangle$, which is given by 
Eqs.~(\ref{inteGam6ti}) and~(\ref{inteGam6totalti}), is the contribution
of the resonance per se. It should be noted that the (total) decay constant 
$\Gamma$ associated with a normalized wave function $\varphi$ is equal 
to one, as it should be, since
the number of events per unit of energy must add up to the
total number of events.

Similarly to $\frac{\rmd \overline{\Gamma}_j(E)}{\rmd E}$ and
$\overline{\Gamma}_j$, we can also define the partial decay constants 
as follows:
\begin{eqnarray}
   \frac{\rmd \Gamma _j(E)}{\rmd E}= \frac{\rmd P_j (E)}{\rmd E}=
       \frac{1}{(E-\er)^2+(\gr/2)^2} |\langle E,j|V \gs|^2 \, , \qquad
         j=1,2  ,
         \label{inteGam66dm} \\ [1ex]
           \Gamma _j = P_j = \int_0^{\infty} \rmd E \
       \frac{1}{(E-\er)^2+(\gr/2)^2} |\langle E,j|V \gs|^2 \, , \qquad
          j=1,2 ,
         \label{inteGam6total6dm}
\end{eqnarray}
where $P_j$ is the probability that the resonance decays into mode $j$. It
follows from Eqs.~(\ref{inteGam66}), (\ref{inteGam6total6}), 
(\ref{inteGam66dm}) and (\ref{inteGam6total6dm}) that
\begin{eqnarray}
   & \frac{\rmd \Gamma _j(E)}{\rmd E} =
       \frac{1}{\gr}
          \frac{\rmd \overline{\Gamma} _j(E)}{\rmd E}  \, , \qquad & j=1,2, 
         \label{inteGam66dmr} \\ [1ex]
   &   \Gamma _j =  \frac{\overline{\Gamma}_j}{\gr} \, , \qquad
          & j=1,2 . 
         \label{inteGam6total6dmr}
\end{eqnarray}

The main advantage of the partial decay constants $\Gamma _j$ over the
partial decay widths $\overline{\Gamma}_j$ is that the decay constants
$\Gamma _j$ make the connection between the experimental and the theoretical 
branching fractions more apparent. In order to see this, let $N_j$ be the 
number of decay events into the $j$ mode. Then, the experimental branching 
fractions are defined as
\begin{equation}
      {\cal B} (\text{R}\to j) = \frac{N_j}{N_1+N_2} \, ,  
             \qquad j =1,2 \, .
            \label{brdc}
\end{equation}  
If the experiment is done $N_0=N_1+N_2$ times (or, equivalently, if there 
are $N_0$ copies of the same resonance), the number of decay events 
into mode $j$ is equal to the number of possible decay events multiplied 
by the probability that the resonance decays into mode $j$,
\begin{equation}
         N_j = N_0 P_j = N_0 \Gamma _j \, , \qquad j=1,2 \, ,
       \label{nimdj}
\end{equation}
where we have used Eq.~(\ref{inteGam6total6dm}) in the second
step. Substitution of 
Eq.~(\ref{nimdj}) into Eq.~(\ref{brdc}) yields
\begin{equation}
      {\cal B} (\text{R}\to j) = 
   \frac{\Gamma _j}{\Gamma _1+ \Gamma _2} \, ,  
             \qquad j =1,2 \, .
            \label{brdcd}
\end{equation}  
Because of Eq.~(\ref{inteGam6total6dmr}), the theoretical branching 
fractions of Eq.~(\ref{brdcd}) are the same as those in 
Eq.~(\ref{br}). However, in Eq.~(\ref{br}) the branching fractions are 
expressed as the quotient of two dimensionful quantities, whereas 
in Eqs.~(\ref{brdc}) and~(\ref{brdcd}) they are expressed as the quotient 
of dimensionless quantities. Thus, it seems preferable to express the 
theoretical branching fractions in terms of the (dimensionless) partial 
decay constants $\Gamma _j$ rather than in terms of the partial decay widths 
$\overline{\Gamma}_j$.

\section{Conclusions}
\setcounter{equation}{0}
\label{sec:conclusions}

The resonant (Gamow) states are the natural wave functions of resonances,
and they lead to expressions for the differential, partial, and total decay 
widths of a resonance. In this paper,
we have presented a simple derivation of the expressions for such
decay widths and have shown that, in the approximation that the
resonance is sharp and far from the lower energy threshold, the expressions
bear a close resemblance to the standard
Golden Rule. When the potential is simple, one can
obtain an exact expression for the decay width (as in the example of the
delta-shell potential), although for most potentials one
is bound to use approximations. We have also constructed the branching fractions
for a resonance that has several decay modes. The resulting branching
fractions coincide with the branching fractions obtained by way of the 
Golden Rule only when the resonance is sharp. Thus, when the resonance 
is not sharp, the branching fractions of the present paper can, at 
least in principle, be distinguished from the standard ones based on 
the Golden Rule.

We have also introduced the (dimensionless) partial, and total
decay constants of a resonant state. Using these dimensionless decay 
constants, we can express both the theoretical and experimental branching 
fractions as the ratio of dimensionless quantities. In addition, the 
differential decay constant can be identified with the energy 
distributions of decay events, which are a non-relativistic analog of the
invariant mass distributions of particle physics.

Although the results of the present paper are restricted to the 
non-relativistic case, it is not hard to imagine a relativistic 
extension. Essentially, in the usual expression for the decay 
rate~\cite{PESKIN}, instead of a delta function we should have 
a truncated, relativistic Breit-Wigner lineshape.

\section*{Acknowledgments}
The author would like to thank the participants of the Quantum Fest, Mexico
City, October 2014, especially Rodolfo Id Betan and Gabriel L\'opez-Castro, 
for stimulating discussions. The author acknowledges 
financial support from Ministerio de Ciencia e Innovaci\'on of Spain under 
project TEC2011-24492.

\appendix
\section{Relationship between $\overline{\Gamma}$ and $\gr$}
\setcounter{equation}{0}
\label{app:rla}

When the survival probability is given by 
$p_{\rm s}(\tau )= \rme ^{-\gr \tau /\hbar}$, the initial decay rate and $\gr$ are
related by
\begin{equation}
       \gr = -\hbar \dot{p}_{\rm s}(0) \, .
            \label{gpd}
\end{equation}
However, if the survival probability was 
$p_{\rm s}(\tau )=p_0 \rme ^{-\gr \tau /\hbar}$, the initial decay rate 
and $\gr$ would be related by
\begin{equation}
       \gr p_0 = -\hbar \dot{p}_{\rm s}(0) \, .
          \label{gpd2}
\end{equation}
Thus, only when the probability is given by the exponential function with
no prefactor, does Eq.~(\ref{gpd}) hold. As we are going to see below,
because the probability of Eq.~(\ref{detdpd}) has a prefactor in front
of the exponential, the decay width of Eq.~(\ref{totaldr}) is different
from the $S$-matrix width $\gr$.

Equation~(\ref{detdpd}) can be written as
\begin{equation}
      \frac{\rmd P_{\tau}}{\rmd E}= \rme ^{-\gr \tau /\hbar} 
         |\langle E|\zr \rangle|^{2}  \, .
             \label{detdpdA}
\end{equation}
Thus, the differential decay width~(\ref{diffdr}) can be written as
\begin{equation}
      \frac{\rmd \overline{\Gamma} (E)}{\rmd E} = \gr |\langle E|\zr \rangle|^{2}  \, .
             \label{diffdrA}
\end{equation}
The total decay width is
\begin{equation}
      \overline{\Gamma} = \gr \int_0^\infty \rmd E \ |\langle E|\zr \rangle|^{2} \, .
         \label{totaldrA}
\end{equation}
Thus, $\overline{\Gamma}$ would be equal to $\gr$ if
\begin{equation}
      \int_0^\infty \rmd E \ |\langle E|\zr \rangle|^{2} =1 \, .
         \label{totaldrA0}
\end{equation}
Since
\begin{equation}
      \int_0^\infty \rmd E \ |\langle E|\zr \rangle|^{2} =
       \int_0^\infty \rmd r \ |\langle r|\zr \rangle|^{2} \, ,
         \label{totaldrA1}
\end{equation}
the decay width $\overline{\Gamma}$ would be equal to the $S$-matrix 
width $\gr$ if the absolute 
value squared of the Gamow state $u(r;\zr) = \langle r|\zr \rangle$ was 
normalized to one,
\begin{equation}
        \int _0^\infty \rmd r \ |u(r;\zr)|^2 =1 \, .
\end{equation}
However, since Zeldovich's normalization is (formally) given by
\begin{equation}
        \int _0^\infty \rmd r \, \left[ u(r;\zr) \right] ^2 = 1 \, ,
\end{equation}
the decay width $\overline{\Gamma}$ is in general different from 
the $S$-matrix width $\gr$.

Finally, we would like to note that, in the standard derivation of the Golden
Rule, it is assumed that the transition probability per unit of time (decay
rate) is independent of time. However, the decay rate associated
with the the survival probability of a resonant state,
$\dot{p}_{\rm s}(\tau)= -(\gr/\hbar) \rme ^{-\gr \tau /\hbar}$, is not 
independent of time, unless we restrict ourselves to a zero-order 
approximation. The same holds for any other probability distribution
that follows the exponential decay law.

\section{The analogy with the Golden Rule}
\setcounter{equation}{0}
\label{app:ana}

In this Appendix, we would like to point out a possible explanation for
why the resonant decay widths of Eqs.~(\ref{inteGam6}) and~(\ref{inteGam6total})
bear such a close resemblance to the standard Golden Rule of 
Eqs.~(\ref{usualgr2}) and~(\ref{usualgr3}) in the approximation that 
the resonance is sharp.

The Lippmann-Schwinger equation for ``in'' states can be written as
\begin{equation}
       |E^+\rangle = |E\rangle + \frac{1}{E-H_0+i\epsilon}V|E^+\rangle \, .
\end{equation}
By writing $G_0=\frac{1}{E-H_0+i\epsilon}$, and by successive substitutions,
we obtain the Born expansion of the Lippmann-Schwinger equation,
\begin{equation}
       |E^+\rangle = |E\rangle + G_0V|E\rangle + 
                    (G_0V)^2|E\rangle +\cdots \, .
        \label{bsls}
\end{equation}
In this expansion, $|E\rangle$ is the term that one would
obtain if there was no interaction, i.e., if $V=0$. Thus, one usually
defines the scattered state, 
$|E_{\rm s}\rangle \equiv |E^+\rangle -|E\rangle$, as the state that contains
the actual effect of the interaction. Equation~(\ref{bsls}) can then be
written as
\begin{equation}
       |E_{\rm s}\rangle =  G_0V|E\rangle + 
                    (G_0V)^2|E\rangle +\cdots \, .
        \label{bslsss}
\end{equation}
The first-order approximation to the scattered state is given by,
\begin{equation}
       |E_{\rm s}\rangle \approx G_0V|E\rangle \, .
       \label{efoap}
\end{equation}
On the other hand, the integral equation satisfied by the resonant states, 
Eq.~(\ref{inteGam0}), can be written as
\begin{equation}
    \gs = G_0(\zr )V \gs \, . 
         \label{inteGam0a}
\end{equation} 
When the resonance is sharp, one can approximate the resonant state by
a bound state $|\er \rangle$,
\begin{equation}
    \gs \approx G_0(\er )V |\er\rangle \, . 
         \label{inteGam0aa}
\end{equation} 
The similarity between Eqs.~(\ref{efoap}) and (\ref{inteGam0aa}) may be 
the reason behind the similarity between the standard Golden Rule of 
Eqs.~(\ref{usualgr2})--(\ref{usualgr3}) and the approximate decay widths of 
Eqs.~(\ref{inteGam8})--(\ref{inteGam9}).

Although Eqs.~(\ref{efoap}) and (\ref{inteGam0aa}) are analogous, there is
a major difference between them. The approximation in Eq.~(\ref{efoap}) 
relies on the assumption that the potential $V$ is weak, whereas the 
approximation in Eq.~(\ref{inteGam0aa}) relies on the assumption that
the potential $V$ is strong enough that the resonance can be 
considered nearly a bound state.


\begin{thebibliography}{99}


\bibitem{CASO} Particle Data Group, S.~Eidelman {\it et al.}, Phys.~Lett.~B
{\bf 592}, 1 (2004).


\bibitem{SIRLIN} A.~Sirlin, Phys.~Rev.~Lett.~{\bf 67}, 2127--2130 (1991);
Phys.~Lett.~B~{\bf 267}, 240--242 (1991).

\bibitem{WILLENBROCK} S.~Willenbrock and G.~Valencia, 
Phys.~Lett.~B~{\bf 259}, 373--376 (1991).

\bibitem{STUART} R.G.~Stuart, Phys.~Lett.~B~{\bf 262}, 113--119 (1991).

\bibitem{LEIKE} A.~Leike, T.~Riemann, and J.~Rose, Phys.~Lett.~B~{\bf 273}, 
513--518 (1991).

\bibitem{BAWI} T.~Bhattacharya and S.~Willenbrock,
Phys.~Rev.~D~{\bf 47}, 4022--4027 (1993).

\bibitem{PASSERA} M.~Passera and A.~Sirlin, Phys.~Rev.~Lett.~{\bf 77}, 
4146--4149 (1996).

\bibitem{BERNICHA} A.~Bernicha, G.~Lopez Castro, and J.~Pestieau, 
Phys.~Rev.~D~{\bf 50}, 4454--4461 (1994); 
Nucl.~Phys.~A~{\bf 597}, 623-635 (1996); 

\bibitem{BH} A.~Bohm and N.L.~Harshman, Nucl.~Phys.~B~{\bf 581}, 
91--115   (2000).

\bibitem{GRASSI} P.A.~Grassi, B.A.~Kniehl, A.~Sirlin,
Phys.~Rev.~Lett.~{\bf 86}, 389--392 (2001).

\bibitem{BS} A.~Bohm, Y.~Sato, Phys.~Rev.~D~{\bf 71},
085018-1--22 (2005).

\bibitem{GP} F.~Giacosa, G.~Pagliara, Phys.~Rev.~C~{\bf 76},
065204-1--10 (2007). 

\bibitem{WOLF} D.N.~Pattanayak, E.~Wolf, 
Phys.~Rev.~D~{\bf 13}, 2287--2290 (1976).


\bibitem{MONDRA} E.~Hern\'andez, A.~Mondrag\'on, 
Phys.~Rev.~C~{\bf 29}, 722--738 (1984).

\bibitem{ARCOS} J.M.~Velazquez-Arcos, C.A.~Vargas, J.L.~Fernandez-Chapou, 
A.L.~Salas-Brito, J.~Math.~Phys.~{\bf 49},
103508-1--16 (2008).

\bibitem{JMP13} R.~de la Madrid, J.~Math.~Phys.~{\bf 53}, 102113 (2012);
{\sf arXiv:1210.3570}. 

\bibitem{GOLDBERGER} M.L.~Goldberger, K.M.~Watson, ``Collision Theory,''
John Wiley \& Sons, New York (1964).

\bibitem{MERZBACHER} E.~Merzbacher, ``Quantum Mechanics,'' 3rd Edition, 
John Wiley \& Sons (1998).

\bibitem{BOHM} A.~Bohm, ``Quantum Mechanics: Foundations and Applications,''
3rd Edition, Springer (1993).

\bibitem{FONDA} L.~Fonda, G.C.~Ghirardi, A.~Rimini, 
Rep.~Prog.~Phys.~{\bf 41}, 587--631 (1978).

\bibitem{LAMB} W.E.~Lamb, H.~Fearn, 
Phys.~Rev.~A~{\bf 43}, 2124--2128 (1991).


\bibitem{COSTIN} O.~Costin, A.~Soffer, 
Comm.~Math.~Phys.~{\bf 224}, 133--152 (2001).

\bibitem{VOS} H.~Thyrrestrup, A.~Hartsuiker, 
J.-M.~G\'erard, W.L.~Vos, Optics 
Express~{\bf 21}, 23130--23144 (2013); {\sf arXiv:1301.7612}.

\bibitem{TOBITA} K.~Ishikawa, Y.~Tobita, 
Prog.~Theor.~Exp.~Phys.\ (2013) 073B02; {\sf arXiv:1303.4568.}

\bibitem{BRYANT} P.W.~Bryant, ``Exponential Decay and Fermi's Golden Rule 
from an Uncontrolled Quantum Zeno Effect,'' {\sf arXiv:1309.5347.}

\bibitem{DEBIERRE} V.~Debierre, I.~Goessens, E.~Brainis, T.~Durt,
``Fermi golden rule beyond the Zeno regime,'' 
{\sf arXiv:1501.04474}.

\bibitem{DEBIERRE2} V.~Debierre, T.~Durt, A.~Nicolet, F.~Zolla,
``Spontaneous light emission by atomic Hydrogen: Fermi's golden rule 
without cheating,'' {\sf arXiv:1502.06404}.


\bibitem{WINTER} R.G.~Winter, Phys.~Rev.~{\bf 123}, 1503--1507 (1961).

\bibitem{DICUS} D.A.~Dicus, W.W.~Repko, R.F.~Schwitters, T.M.~Tinsley,
Phys.~Rev.~A~{\bf 65}, 032116-1--5 (2002).

\bibitem{GASTON} G.~Garcia-Calderon, Advances in Quantum Chemistry~{\bf 60}, 
407--455 (2010).

\bibitem{SANTINI1} U.G.~Aglietti, P.M.~Santini, Phys.~Rev.~A~{\bf 89}, 
022111-1--12 (2014); {\sf arXiv:1303.4977}.

\bibitem{SANTINI2} U.G.~Aglietti, P.M.~Santini, ``Geometry of Winter Model,''
{\sf arXiv:1503.02532}.

\bibitem{IJTP} R.~de la Madrid, Int.~J.~Theor.~Phys.~{\bf 42}, 
2441--2460 (2003); {\sf arXiv:quant-ph/0210167}. 

\bibitem{ATLAS} The ATLAS Collaboration, Phys.~Lett.~B~{\bf 716},
1--29 (2012); {\sf arXiv:1207.7214}

\bibitem{CMS} The CMS Collaboration, Phys.~Lett.~B~{\bf 716}, 30--61 (2012);
{\sf arXiv:1207.7235}.

\bibitem{LHCb} The LHCb Collaboration, Phys.~Rev.~Lett.~{\bf 113}, 
172001-1--9 (2014); {\sf arXiv:1407.5873}.

\bibitem{CDF} The CDF Collaboration, Phys.~Rev.~D~{\bf 90}, 
091101-1--8 (2014); {\sf arXiv:1409.4906}.

\bibitem{BESIII} The BESIII Collaboration, Phys.~Rev.~D~{\bf 89}, 
112006-1--8 (2014); {\sf arXiv:1405.1571}.

\bibitem{BELLE} The BELLE Collaboration, Phys.~Rev.~D~{\bf 90}, 
112008-1--9 (2014); {\sf arXiv:1409.7644}. 

\bibitem{D0} The D0 Collaboration, Phys.~Rev.~Lett~{\bf 114}, 
062001-1--7 (2015); {\sf arXiv:1410.1568}.

\bibitem{BABAR} The BABAR Collaboration, Phys.~Rev.~D~{\bf 91}, 
012003-1--12 (2015); {\sf arXiv:1407.7244}.

\bibitem{ALICE} The ALICE Collaboration, Phys.~Lett.~B~{\bf 740}, 
105--117 (2015); {\sf arXiv:1410.2234}.


\bibitem{PESKIN} M.E.~Peskin, D.V.~Schroeder, ``An Introduction to Quantum
Field Theory,'' Addison-Wesley (1995).



\end{thebibliography}
\end{document}